Linking Europa's plume activity to tides, tectonics, and liquid water

Alyssa Rose Rhoden*[1,2], Terry A. Hurford[1], Lorenz Roth[3,4], and Kurt Retherford[3]

1 – NASA Goddard Space Flight Center, Code 693, Greenbelt, MD 20771
2 – Johns Hopkins University Applied Physics Laboratory, Laurel, MD 20723
3 – Southwest Research Institute, San Antonio, TX 78238
4 – School of Electrical Engineering, Royal Institute of Technology, Stockholm, Sweden

* - Previously published under the surname Sarid

Highlights:

- Tidal models with spin pole precession are consistent with Europa's plume variability
- Our analysis indicates that either active fractures or volatile sources are rare
- Europa's plumes are likely more variable in time and location than Enceladus' plumes

Manuscript pages: 24
Figures: 3 in main text, 5 in SOM
Tables: 2

Tidal-tectonic plume sources on Europa

Corresponding author:	Alyssa Rhoden
　　　　　　　　　　　　Johns Hopkins University Applied Physics Lab
　　　　　　　　　　　　Laurel, MD 20723
　　　　　　　　　　　　Alyssa.Rhoden@jhuapl.edu

Abstract:

Much of the geologic activity preserved on Europa's icy surface has been attributed to tidal deformation, mainly due to Europa's eccentric orbit. Although the surface is geologically young (30 – 80 Myr), there is little information as to whether tidally-driven surface processes are ongoing. However, a recent detection of water vapor near Europa's south pole suggests that it may be geologically active. Initial observations indicated that Europa's plume eruptions are time-variable and may be linked to its tidal cycle. Saturn's moon, Enceladus, which shares many similar traits with Europa, displays tidally-modulated plume eruptions, which bolstered this interpretation. However, additional observations of Europa at the same time in its orbit failed to yield a plume detection, casting doubt on the tidal control hypothesis. The purpose of this study is to analyze the timing of plume eruptions within the context of Europa's tidal cycle to determine whether such a link exists and examine the inferred similarities and differences between plume activity on Europa and Enceladus. To do this, we determine the locations and orientations of hypothetical tidally-driven fractures that best match the temporal variability of the plumes observed at Europa. Specifically, we identify model faults that are in tension at the time in Europa's orbit when a plume was detected and in compression at times when the plume was not detected. We find that tidal stress driven solely by eccentricity is incompatible with the observations unless additional mechanisms are controlling the eruption timing or restricting the longevity of the plumes. The addition of obliquity tides, and corresponding precession of the spin pole, can generate a number of model faults that are consistent with the pattern of plume detections. The locations and orientations of these hypothetical source fractures are robust across a broad range of precession rates and spin pole directions. Analysis of the stress variations across the fractures suggests that the plumes would be best observed earlier in the orbit (true anomaly ~120°). Our results indicate that Europa's plumes, if confirmed, differ in many respects from the Enceladean plumes and that either active fractures or volatile sources are rare.

1. Introduction

Europa's surface records a rich history of tectonic activity, including linked arcuate fractures called cycloids and straighter fractures called lineaments. Dilation, compression, and shear motions are also observed along many of these fractures, suggesting that fractures can be subjected to a variety of post-formation processes (for a review, see Kattenhorn and Hurford, 2009). Tectonic activity on Europa has been largely attributed to daily variations in tidal stress induced by Europa's eccentric orbit (e.g. Greenberg et al., 1998). Eccentricity causes Jupiter's location relative to Europa to change throughout each Europan day, leading to global-scale deformation and tidal stress. Lineaments and cycloids are both hypothesized to form mainly in response to tensile tidal stress, while strike-slip offsets have been linked to tidal shear stress (e.g. Greenberg et al., 1998; Rhoden et al., 2012).

Earth-based radar measurements show that Europa's obliquity (i.e. tilt of the spin pole) is also non-zero (Margot et al., 2013), likely forced by torques from Jupiter's other large moons (Bills et al, 2009; Baland et al., 2012). The maximum obliquity predicted by gravitational models is 0.1° (Bills et al, 2009; Baland et al., 2012), but the value depends sensitively on the interior structure assumed for Europa. For example, Bills et al. (2009) assumed that Europa is a homogeneous sphere. The models also predict that Europa's obliquity will change in magnitude periodically over 10 to 1000-year timescales. In addition, Europa's spin pole should precess at a rate of 0.3° to 3° per day (Bills et al., 2009). Obliquity and spin pole precession augment the stress field generated by eccentricity (e.g. Fig. 3 of Rhoden et al., 2010). The timescales for variation in the magnitude of the obliquity and the spin pole direction are short enough to be geologically-relevant.

Detailed analyses of tectonic features on Europa indicate that obliquity was a factor in their formation (Rhoden et al., 2010; Rhoden et al., 2012; Rhoden and Hurford, 2013), although they correspond to a somewhat higher obliquity than the gravitational models predict (between 0.2° and 1° depending on feature type). The tectonic record further supports the idea that several precession periods have occurred within the surface age and that there is a stronger signal of tidal stress change due to spin pole precession than non-synchronous rotation of the ice shell (Rhoden and Hurford, 2013). However, the absolute ages of tectonic features are unknown; interpretations of fractures are also compatible with a much slower precession rate than the gravitational models suggest.

Europa was clearly geologically active in the past, but whether it is still active today has remained a topic of much debate. However, recent observations, made with the Space Telescope Imaging Spectrograph (STIS) of the Hubble Space Telescope (HST), revealed H and O emissions near Europa's south pole that are best explained by two ~200-km high plumes of water vapor (Roth et al., 2014a). The plumes were identified in observations taken when Europa was near apocenter; no plumes were identified in two earlier observations when Europa was closer to pericenter. The apparent correlation with Europa's orbital position (i.e. true anomaly) was interpreted as possible evidence of a tidal origin for the plumes (Roth et al., 2014a). Similarly, active plumes emanating from fractures near the south pole of Saturn's moon, Enceladus (Porco et al., 2006; Spencer et al., 2006), have been shown to vary in intensity with its orbital position (Hedman et al., 2013) and have been linked to eccentricity tides (Hurford et al., 2007; Nimmo et al., 2007; Hurford et al., 2012; Porco et al., 2014; Nimmo et al., 2014). However, tidal-control of Europa's putative plumes is still uncertain, especially after follow-up HST/STIS observations taken when Europa was once again near apocenter failed to yield a repeat detection (Roth et al., 2014b).

It is important to note that non-detections do not necessarily imply a change in $H_2O$ plume abundance. Several effects contribute to detectability, such as the highly variable local plasma environment and the observing geometry (Roth et al. 2014b). However, to examine the potential role of tidal stress in controlling plume eruptions on Europa, we will begin with the assumption that non-detections indicate a lack of plumes.

The daily pattern of stress change on a fault due to Europa's eccentric orbit repeats exactly from one orbit to the next. Although the initial plume variability reported by Roth et al. (2014a) appeared consistent with eccentricity-driven tidal stress, the non-detection in Jan 2014 all but rules out an eccentricity-only model for controlling plume eruptions on Europa. An alternative explanation is that an additional mechanism acts to change the stress state on a fault even at the same true anomaly. Precession of Europa's tilted spin pole will cause the daily variation in tidal stress on a fault to slowly change over successive orbits. Because the five sets of HST/STIS observations obtained, to date, span a period of 14 years, the precession rates indicated by gravitational models would lead to a different spin pole direction in each of the observations.

We assess the conditions under which daily-varying tidal stress would be consistent with the plume detection pattern identified in the five available HST/STIS observations of Europa. Lessons learned from Enceladus are described in Section 2. Our tidal stress calculations, the stress conditions we assume for plume generation, and the different rotation states we consider are described in Section 3. Under these assumptions, we find that precession of Europa's slightly tilted spin pole can alter the stress state between observations, even when Europa is at a similar orbital position, such that tidal stresses are compatible with the pattern of plume detections and non-detections. The locations and orientations of the model faults that are most consistent with the observations, along with the inferred constraints on Europa's rotation state, are described in Section 4. Implications of our findings on the distribution of plume sources, future observations that can further refine (or refute) the tidal-control hypothesis, and comparisons with Enceladean plumes are discussed in Section 5.

2. Plumes on Enceladus

The *Cassini* spacecraft identified active and continuous venting of material emanating from the south pole of Enceladus (Waite et al., 2006; Hansen et al., 2006; Dougherty et al., 2006; Tokar et al., 2006; Porco et al., 2006; Spencer et al., 2006), which has now been shown to be the source of Saturn's E-ring (Kempf et al., 2010). The material forms a broad plume that has erupted material throughout the *Cassini* observations (>10 yrs) as well as in individual, intermittent jets (Spitale and Porco, 2007, Porco et al., 2014). Continued observations using the *Cassini* VIMS instrument revealed that the intensity of the plume eruption varies with Enceladus' orbital position (Hedman et al., 2013). The main source of both the broad plume and the jets is a set of parallel fractures (Spitale and Porco, 2007), dubbed Tiger Stripes, which are also anomalously warm (Spencer et al., 2006; Spencer and Nimmo, 2013).

Like Europa, Enceladus maintains a mean motion resonance with neighboring

Dione, which forces its eccentricity to 0.0047 (Peale, 1976). Hence, Enceladus experiences cyclic tidal stresses on the timescale of its 1.37-day orbit around Saturn. Tidal stress normal to the Tiger Stripes has been linked to plume eruptions, although the relationship between tides and individual jet activity may be more complex than the behavior of the broader plume (Hurford et al., 2007, 2012; Spitale and Porco, 2007; Porco et al., 2014). Tidal shear stress has been proposed to explain both heating and eruptions along the Tiger Stripes (Nimmo et al., 2007), but further analysis of plume activity suggests that the eruption timing is more consistent with normal stress (Porco et al., 2014). These interpretations suggest that the Tiger Stripe fractures mark the locations of conduits to a volatile reservoir that is opened and closed under the influence of tides.

Enceladus' shape and gravity field are most consistent with a regional sea near Enceladus' south pole, which sits beneath an ice shell 10s of km thick (Collins and Goodman, 2007; Iess et al., 2014). However, a global ocean that is considerably thicker/shallower at the south pole cannot be ruled out and may be more consistent with older fractures (Patthoff and Kattenhorn, 2011). In either case, the south polar sea is the most plausible source for the plume material, although the exact mechanism by which fractures interact with the liquid reservoir is unknown. The Tiger Stripe fractures may penetrate through the entire ice shell and tap the ocean directly, or there may be a more complex process of material transport that brings liquid water from the ocean to the shallow subsurface (see review by Spencer and Nimmo, 2013). Because tidal stresses reach a maximum of ~100 kPa, overburden stresses should dominate deeper than a few hundred meters and restrict opening of the Tiger Stripe fractures. However, hundreds of observations of plume activity at Enceladus demonstrate that a viable mechanism does exist for connecting fractures at the surface with liquid water beneath an ice shell 10's of km thick.

There is now evidence that both Enceladus and Europa have icy surfaces and subsurface reservoirs of liquid water, display tidal-tectonic activity driven by forced orbital eccentricity, and actively vent water in the form of large plumes. Comparing the behavior of plumes on Europa and Enceladus can thus provide insight into the process of geyser activity and tidal-tectonics on icy satellites. However, before we assume that lessons learned at Enceladus can be applied to Europa, we must characterize any differences in the activity of these two bodies (see Section 5.3).

3. Methodology

3.1 Comparing model predictions with plume observations at Europa

One set of HST observations of Europa's UV emission revealed Lyman-α emission consistent with two plumes of water vapor (Roth et al., 2014a) in the southern hemisphere; four other sets of HST observations found no evidence of plumes. It is possible that non-detections are caused by lower abundances of plume material rather than complete absence of a plume. A putative set of persistent plumes would be limited to factors of 2-4 times lower abundances, constrained by the brightness upper limits (Roth et al., 2014a,b). Such variation is observed for Enceladus' plumes (Hedman et al., 2013). Details of all the HST observations presented by Roth et al. (2014a,b) are shown in Table 1. The most likely locations of the plume sources were between latitudes 50°S and 80°S and longitudes 120°W and 250°W, with a narrower range of longitudes for more northern latitudes. For reference, 180°W is the anti-Jupiter point.

In both the Dec 2012 and Jan 2014 observing windows, Europa moved from an orbital position of 191° to 218°, which made up 90% of each observing window, yet the plume was only detected in Dec 2012. For eccentricity-driven tidal stress to independently explain the observations, the plume must have been active only during the 5% of the Dec 2012 observing window when Europa was at a true anomaly not observed in Jan 2014. In that case, the tensile stress across the source fault must have been nearly zero as Europa moved from a true anomaly of 189° to 191° such that it would have been in tension at the beginning of the Dec 2012 observing window and in compression by the beginning of the Jan 2014 observing window. This model would imply that the plumes remained active until just before the source fractures went into compression, that the plumes were observed at exactly this transition time, and that the transition occurred within the first 5% of the Dec 2012 observing window, such that Europa's true anomaly did not overlap with the range of true anomalies from Jan 2014 – an extremely fortuitous observation! Although we cannot rule out an eccentricity-only model for tidal control of Europa's plumes, we focus our analysis on eccentricity-driven tidal stress modulated by Europa's non-zero obliquity.

To compare the plume observations to tidal model predictions, we first calculate the daily tidal stress normal to hypothetical fractures on a latitude-longitude grid that covers all longitudes, in increments of 15°, and latitudes from 45°S to 85°S, in increments of 5°. At each location, we test fracture azimuths from 0° (N) to 180° (S) using an increment of 1°. This results in 38,880 location-azimuth pairs, which represent hypothetical source fractures. For a given location-azimuth pair, the corresponding fracture would experience varying tidal stress throughout each 85-hour Europan day. We then use Europa's orbital position at the midpoint of each 7-hour HST observation window (Table 1) to determine the stress on each fault at the time of the observations. The choice of when within the window to make our stress calculations has a small effect on the azimuths of the resulting candidate fractures (see SOM), but it does not change our overall conclusions about the potential role of obliquity and precession in controlling plume eruptions.

We expect some delay between the eruption of material at Europa's surface and the appearance of plumes at heights detectable with HST. Roth et al. (2014a) estimated the velocity of plume particles to be ~560 m/s, enabling plume particles to reach 200 km in about 10 minutes. During this time, Europa would move <1° along its orbital path. Hence, the appearance of plumes should depend on the stress state slightly earlier in Europa's orbit than the plumes were observed. Because the tidal stresses do not change significantly over increments of ~1°, the delay in detectability would have a negligible effect on our results.

Although Europa's non-zero obliquity has been confirmed (Margot et al., 2013), data is still being analyzed to determine the exact value of the obliquity, the direction of the spin pole, and the spin pole precession rate. Hence, for this analysis, we rely on studies that model the effects of mutual gravitational interactions of the Galilean satellites on Europa's spin pole. These studies indicate that Europa's obliquity should be between 0.033 and 0.1°, depending on assumptions about Europa's internal structure (Bills et al., 2009; Baland et al., 2012), although a larger obliquity cannot be ruled out (Bills et al., 2009). The corresponding precession rate should be between 0.3 and 3° per day (Bills et al., 2009). Studies of tidal-tectonic fractures on Europa suggest a larger obliquity of order 1° (Rhoden et al., 2010; Rhoden et al., 2012; Rhoden and Hurford, 2013), which may indicate a past period of high obliquity (e.g. due to an impact) or that an additional mechanism is forcing the obliquity.

For the present analysis, we use an obliquity of 0.1°, which is at the upper end of the gravitational modeling results. A smaller obliquity would cause a decrease in the magnitude of the tidal stress (Rhoden et al., 2010; Jara-Orue and Vermeersen, 2011) and, more importantly, narrow the range of latitudes for which spin pole precession can significantly alter tidal stress patterns (Hurford et al., 2009). There are no constraints on Europa's spin pole direction (SPD) during the HST observations. Hence, we test a number of SPDs, with particular focus on explaining the detection and non-detection at similar orbital positions. Plumes were detected in the December 2012 observation, during which Europa moved from 189° to 218° in its orbit (Roth et al., 2014a), However, plumes were not detected in January 2014, when Europa was at similar orbital positions: 191° - 221° (Roth et al., 2014b). Because the precession rate is only loosely constrained, we test rates that are consistent with the gravitational modeling results and maximize the difference in stress between these two observing windows (0.15° to 0.5° per day). We also analyze cases in which the spin pole direction remains constant for all observations (i.e. nearly zero precession rate).

We hypothesize that plume eruptions occur when a preexisting fracture experiences tensile tidal stress, similar to the mechanism proposed for plumes on Enceladus (Hurford et al., 2007, 2012; Nimmo et al., 2014; Porco et al., 2014). Tensile stress across the fracture could open it, creating a conduit for subsurface plume material. Likewise, the fracture (and conduit) would close when the tidal stress becomes compressive. This model is certainly an oversimplification of the eruption process. For example, there is likely a delay between compression closing a source fracture and the plume no longer being detectable. Hence, faults that were just entering a compressive stress state in Jan 2014 may be less compatible with a non-detection at that time than faults that were in compression for a larger portion of the orbit. However, our lack of knowledge about the dynamics of the plumes warrants a simplified model. To determine the conditions under which our hypothesis is compatible with the plume observations, we identify those fractures for which the tidal stress normal to the fracture would have been tensile at the midpoint of the Dec 2012 observing window, when the plume was detected, and compressive at all other observing midpoints. We refer to hypothetical fractures that meet this requirement as "candidate fractures".

3.2 Calculating tidal stress

To calculate tidal stress on Europa, we assume a thin, elastic, outer shell that is mechanically decoupled from the interior, a reasonable assumption due to Europa's global, subsurface ocean. The principal surface stresses generated by the primary tide are given by (e.g., Melosh 1977; Melosh, 1980):

(1a) $\sigma_1 = C (5 + 3 \cos 2\delta_P)$
(1b) $\sigma_2 = -C (1 - 9 \cos 2\delta_P)$

where $C = 3h_2 M\mu(1 + \nu)/ 8\pi\rho a^3(5 + \nu)$ and $h_2$ is the tidal Love number, $M$ is Jupiter's mass, $\mu$ is the shear modulus, $\nu$ is Poisson's ratio, $\rho$ is the average density, $a$ is the distance between Jupiter and Europa, and $\delta_P$ is the angular distance from a point on Europa's surface to the primary tidal bulge. The $\sigma_1$ stress is directed radially from the tidal bulges and the $\sigma_2$ stress is perpendicular to the $\sigma_1$ stress. All the stress calculations contain a factor of C, so its value will only change stress magnitudes, not the sign of the stress (tensile or compressive) that forms the basis of this analysis.

Europa's eccentricity causes the tidal bulges to librate in longitude; obliquity predominantly causes a latitudinal libration. We use spherical trigonometry to calculate the time-varying location of the tidal bulge, which depends on the spin pole direction and the true anomaly. Since the locations of the tidal bulges are changing, the angular distance to the bulge, $\delta$, also changes. The tidal stress equations can thus be modified to account for eccentricity and obliquity (see also Hurford et al. 2009).

(2a) $\sigma_{1*} = C (1 - e \cos n)^{-3} (5 + 3 \cos 2\delta (\lambda,\omega))$

(2b)     $\sigma_{2e} = -C (1 - e \cos n)^{-3} (1 - 9 \cos 2\delta (\lambda,\omega))$

with the important distinction that the value of $\delta$ now depends on the bulge colatitude, $\lambda$, and the bulge longitude, $\omega$, as follows:

(3a)     $\lambda = \pi/2 - \varepsilon \sin(n + \varphi)$

(3b)     $\omega = -2e \sin n$

In the above equations, e is eccentricity, $\varepsilon$ is obliquity, $\varphi$ is the spin pole direction (SPD), and $n$ is the true anomaly (our $n$ is equivalent to $f$ in Roth et al., 2014a,b). When Europa is at pericenter and the spin pole is pointing toward Jupiter, the SPD is defined as $\varphi = 90°$; SPD increases clockwise as viewed from Europa's north pole. For models that include spin pole precession, we compute the SPD for each observation based on the assumed precession rate, the assumed SPD for the Dec 2012 observation, and the number of days between observations.

Equations 2a and 2b provide the total tidal stress due to eccentricity and obliquity. To isolate the diurnal tide, we rotate the total tidal stress (Eqs. 2a and 2b) and the primary tidal stress (Eqs. 1a and 1b) to a common coordinate system and subtract out the primary tidal stress. In the new coordinate system, the diurnal stress has two normal stress components ($\sigma_\alpha$ and $\sigma_\beta$) and a shear stress component ($\sigma_{\alpha\beta}$).

Finally, we decompose the diurnal tidal stresses into normal ($\sigma$) and shear ($\tau$) components relative to a hypothetical fracture's orientation, where $z$ is the azimuth of the fracture measured clockwise from north.

(4a)     $\sigma = 0.5 (\sigma_\alpha + \sigma_\beta) + 0.5 (\sigma_\alpha - \sigma_\beta) \cos(2z) + \sigma_{\alpha\beta} \sin(2z)$

(4b)     $\tau = -0.5 (\sigma_\alpha - \sigma_\beta) \sin(2z) + \sigma_{\alpha\beta} \cos(2z)$

More sophisticated approaches for calculating tidal stress are also available (e.g. Wahr et al., 2009; Jara-Orue and Vermeersen, 2011). These approaches take, as inputs, characteristics of Europa's internal structure including the thicknesses and viscosities of internal layers. Use of these techniques shows that diurnal tidal stress in thick shells (e.g. 30 km) will differ in magnitude by, at most, 3% from stress derived using the thin shell approximation (Eqs. 1-3). The timing of stress change at a particular location on Europa can also differ between the two approaches (Jara-Orue and Vermeersen, 2011). For longer period stresses (e.g. non-synchronous rotation), the difference between techniques has a larger impact on the results. We adopt the thin shell model for our main analysis of different tidal models because (1) the effects of stress relaxation in the ice shell are unlikely to significantly affect stress changes over a tidal cycle, (2) we are not attempting to constrain the interior with our analysis, and (3) there are significant uncertainties in Europa's interior structure and rheology.

However, to determine the robustness of our results, we apply the approach of Jara-Orue and Vermeersen (2011) to compute stresses in a viscoelastic shell using one of our tidal models. This approach utilizes the propagator matrix method (Love, 1927; Alterman et al., 1959; Takeuchi et al., 1962; Peltier, 1974; Sabadini and Vermeersen, 2004) to compute the loading response of a series of laterally homogeneous incompressible material layers within the Laplace domain. For details of the approach, we point the reader to Jara-Orue and Vermeersen (2011) and Rhoden et al. (2014), in which we apply this technique to Pluto's moon, Charon. Here, we use this method only to determine the extent to which the resulting predictions of candidate fractures differ from those derived using the thin shell approximation.

Following Jara-Orue and Vermeersen (2011), we assume that Europa' interior includes a core, mantle, liquid ocean, and 30-km ice shell that is approximated as a linear Maxwell viscoelastic material. The ice shell is split into a 5-km brittle upper layer with a viscosity of $10^{21}$ Pa*s and a 25-km ductile lower layer with a viscosity of either $10^{17}$ Pa*s (the "dissipative" case) or $10^{19}$ Pa*s (the "elastic" case). For all other parameters (e.g. eccentricity), we use the same values as in our thin shell calculations. We find that the locations and azimuths of candidate fractures identified using this tidal model are nearly identical whether we apply the thin shell approximation or use the viscoelastic approach for an elastic 30-km shell. Hence, the simplified thin shell approach is sufficient for this analysis.

4. Results

Assuming either zero obliquity or a constant spin pole direction means that the stress on a given fault does not change from orbit to orbit. Hence, there are similar challenges in fitting the observations in both cases. We find that 0.8% of the location-azimuth pairs we tested (as described in section 3.1) can meet our stress requirements using these models, and that the stress on faults in Dec 2012 had to be nearly zero in order to be compressive by the start of the Jan 2014 observing window. Unlike the precession models (described below), we used the starting points of the observing windows for this analysis in order to avoid the overlap in true anomaly between the Dec 2012 and Jan 2014 observing windows.

When we adopt a faster precession rate, the stress on a particular fault can change from one orbit to the next due to the change in spin pole direction. In Table 2, we list the different precession rates we tested, the specific SPDs we selected for the Dec 2012 and Jan 2014 observations, and the results for each case. The largest difference in the variation of daily stress for faults in the southern, sub-Jovian hemisphere corresponds to a change in spin pole direction from 270° to 90° (i.e. a spin pole tilted away from Jupiter when Europa is at pericenter to a spin pole tilted toward Jupiter when Europa is at pericenter). Reducing the precession rate reduces the number of candidate faults that meet our stress criteria and the magnitude of the stress on the fault during the Dec 2012 observation. When testing different precession rates, we used two approaches in order to determine the sensitivity of the results to each parameter. In one case, we used an SPD of 270° for the Dec 2012 observation and varied the SPD for the Jan 2014 observation to reduce the amount of precession between observations. In the other case, we narrowed the range of SPD change (centered on an SPD of 0°) by selecting a different SPD for both the Dec 2012 and Jan 2014 observations.

Assuming an SPD of 270° in the Dec 2012 observation and an SPD of 90° in the Jan 2014 observation (a precession rate of 0.464°/day or odd multiples of it), we find that 1.7% of the location-azimuth pairs we tested qualify as candidate fractures. Furthermore, the stress in Dec 2012 is no longer limited to very small values, unlike the no-obliquity or constant SPD models. In Figure 2, we show the variation in daily stress on one of these candidate faults due to the change in SPD from one observing time to the next. Stress throughout one orbit is shown for each observing window, which corresponds to its specific SPD, derived based on the assumed precession rate. Red, filled symbols mark Europa's orbital position at the midpoint of each observing window, while the red line indicates the full range of positions covered during each 7-hr window. The change in spin pole direction between the Dec 2012 and Jan 2014 observations causes the stress curve to shift earlier in the orbit (to the left in Fig. 1), allowing the stress at the midpoint to switch from tensile to compressive despite having nearly the same orbital position and without requiring that the stress be nearly zero when the plume was observed. This candidate fracture is the one that experiences the highest tensile stress at the Dec 2012 midpoint. Higher stress allows the fracture to penetrate deeper and remain in tension longer, which could promote eruptions and detection. As shown in Fig. 1, this particular fracture would have been in tension throughout the entire Dec 2012 observing window, during which time the plume was detected, and in compression over the majority of the Jan 2014 observing window.

      The distribution of candidate fractures is fairly robust across the range of precession parameters we tested. Figure 3 shows the azimuths of the candidate fractures (black lines) embedded in latitude-longitude bubbles for the precession model that produced the largest number of candidate fractures. The azimuths tend to cluster, creating wedges in each bubble. The largest wedges are at the most northern latitudes we tested (45°S) because the effects of obliquity on the stress field are most pronounced closer to the equator. We selected SPDs specifically to identify faults in this hemisphere due to the observational constraints on the plume locations from Roth et al. (2014a); different SPD values would shift the pattern in longitude. Candidate fractures in the largest clusters (e.g. 45°S, 150°W) are oriented NW-SE to N-S. The azimuth distributions gradually narrow and the fracture orientations trend more E-W with decreasing latitude and a change in longitude away from the sub-Jupiter/trailing hemisphere. The distributions for slower precession rates are shown in the SOM. In those cases, the sizes of some wedges change slightly from what is shown in Figure 3. However, the general locations and orientations of candidate faults remain the same.

5. Discussion

5.1. Tidal control of plumes on Europa

The overlap in Europa's true anomaly between the Dec 2012 and Jan 2014 observing windows poses a particular challenge to a tidal model that includes stress only from eccentricity. An alternative scenario is that precession of Europa's spin pole plays a role in controlling the eruptions. If the spin pole direction changed by at least 60° between the Dec 2012 and Jan 2014 observations, the stress on a subset of faults would have changed from tensile to compressive, respectively, even though Europa's orbital position was the same during both observations.

We find the largest number of candidate fractures and the highest tensile stress magnitudes during the Dec 2012 observation using an SPD of 270° for the Dec 2012 observations and an SPD of 90° for the Jan 2014 observations, a difference of 180° in 388 days. These SPDs correspond to a precession rate of 0.464°/day, which is well within the rates derived from gravitational models (Bills et al., 2009). Reducing the precession rate reduces the number of candidate fractures and the maximum tensile stress across the fractures. With a precession rate of 0.155°/day, the lowest rate we tested, the results approach those of the constant SPD cases, in which the tidal stress patterns on faults remain constant over all of the HST observations. Due to the cyclic nature of spin pole precession, increasing the precession rate above 0.464 will similarly reduce the number of candidate fractures.

For this analysis, we have assumed that <1 precession period occurred between the Dec 2012 and Jan 2014 observations. If we assumed some arbitrary number of additional precession periods, the precession rates would be higher by 0.93 °/day per period, but the peak in the number of candidate faults would still occur when the SPD was 270° in Dec 2012 and 90° in Jan 2014. Changing the precession rate slightly changes the distribution of candidate fractures because the rate determines the SPD we assume for each observation time, thereby changing the orientations and locations of fractures that experience tension only during the Dec 2012 observation. These subtle changes are unlikely to have much diagnostic power.

We made the simplifying assumption that plumes are on when the source fractures are in tension and off when they are in compression. This approach was used to predict the periodic nature of the Enceladean plumes (Hurford et al., 2007) and roughly correlates with the timing of peak emission within Enceladus' orbit (Hedman et al., 2013). However, this simple model has not reproduced the detailed temporal variability of Enceladus' plumes (Hurford et al., 2012; Porco et al., 2014). Additional characteristics of the stress field, such as the magnitude of the tidal stress or viscous relaxation within the ice shell, or the interactions between fractures and the liquid water reservoir likely contribute to the overall pattern of eruptive output. Furthermore, applying this simplifying assumption means that the candidate fractures we identify are likely only a subset of plausible candidates. For example, fractures experiencing very low tensile stress would probably be consistent with non-detection of plumes because the penetration depth of the fractures would be so low. In our models, we require that the tidal stress be compressive at the midpoints of all of the observing windows associated with non-detections. With only one detection of plumes on Europa, the simplified approach is sufficient to draw broad conclusions about the likelihood of tidal control of plumes.

Additional plume observations can further refine, or refute, the tidal model derived here, especially if the observations are timed to provide the most diagnostic tests. Follow up observations that aim to replicate the orbital and rotational conditions during the Dec 2012 detection would be challenging given that the plausible range of spin pole periods is 2 to 6 years, based on our analysis. However, there is a more straightforward test. All of the candidate fractures we have identified must go from tension to compression just after Europa passes through apocenter in order to meet the stress constraints. Due to the periodic nature of tidal stress, these fractures should experience a peak in tensile stress when Europa is at about 120° in its orbit (i.e. ~90° earlier), as illustrated in Fig. 2. The spin pole direction will shift the stress curve in time, but not enough to put these fractures into compression at an orbital position of 120°. If the plumes are active when the source faults are in tension, as we have assumed here, then this orbital position would be the best opportunity to detect them.

If the plumes are not detected when Europa is at 120°, it would strongly suggest that the plumes are not periodic on the timescale of years. In that case, either tides played no role in the observed eruption, the eruption was short-lived such that the periodicity was not captured in the observations, or the tidal effects controlling the plume are being modulated by an additional process that operates on a longer timescale. For example, a short-lived periodic plume might be the result of a small liquid reservoir that was depleted between observations; a modulating process could be the time to refill a liquid reservoir after an eruption.

5.2 Comparison with other tidally-controlled activity on Europa

We find that the pattern of plume detections, if tidally-controlled, is most consistent with a rapidly precessing spin pole (period of order years) because a change in spin pole direction can generate different stress conditions on a fault even when Europa is at the same orbital position. This result is consistent with gravitational models, but how does it compare with the rates inferred from modeling of Europa's tidal-tectonic features?

The best fits to six observed cycloids, whose shapes depend sensitively on spatial and temporal variations in tidal stress, indicate different spin pole directions for each feature (Rhoden et al., 2010), which implies that some precession occurred between the formation of each cycloid. Allowing the spin pole to precess within the formation timescale of an individual cycloid did not significantly improve fits. Assuming the hypothesized formation rate of one arc per day, the lack of precession signal within individual cycloids suggests rates of 0.5° per day or less, consistent with our plume analysis.

Global strike-slip fault patterns and the azimuth distribution of lineaments are best explained if the majority of the features formed at an SPD of 120° (Rhoden et al., 2012, Rhoden and Hurford, 2013), although both populations also contain features formed at other SPDs. The combination of an excess of features consistent with one SPD and the breadth of lineament azimuths overall suggests at least one precession period elapsed during the formation of the most recent features, but not many. The faster the spin pole precesses, the harder it is to preserve the fault patterns associated with any one SPD. Hence, similar to the results for cycloids, this implies a precession rate that is fast relative to the surface age but slow with respect to the formation timescale of individual features.

Unfortunately, the formation timescales are not well constrained.

If the precession rate is as fast as the gravitational models (and our plume analysis) would suggest, it seems unlikely that the patterns associated with individual SPDs would be preserved in the tectonic record. However, the effects of overprinting have not been quantified, so we cannot rule out rapid precession based on the tectonic record. It is also possible that Europa's rotation state has changed with time. For example, the tectonic models indicate a larger obliquity (of order 1°) than the gravitational models (e.g. Baland et al., 2012) or the preliminary results of radar measurements (Margot et al., 2013). Perhaps the precession rate has also varied with time as the magnitude of the obliquity has changed. Measurements of Europa's current spin pole direction and precession rate will help differentiate between these scenarios. A very slow precession rate is still compatible with a tidal-control model for the plumes, but it requires a very fortuitous observation to explain the detection in Dec 2012.

5.3 Additional implications and comparison with Enceladean plumes

The candidate faults we identify are the only ones that would have experienced tension only during the Dec 2012 observation. However, it is important to note that stresses on these particular faults are not otherwise different from other faults on Europa. Of the 38,880 location-azimuth pairs we tested, 64% of the corresponding faults would have been in tension during the Dec 2012 observation, and the tensile stress across 52% of them was at least as high as on any of our candidate fractures. Furthermore, 63% of faults would have been in tension during the Jan 2014 observation when no plume was detected. If tension across a fault was the only requirement to produce a plume, there should have been numerous plumes visible both in Dec 2012 and Jan 2014. These results imply that some additional mechanism, beyond tidal stress, is controlling large plume eruptions on Europa.

At Enceladus, plumes are only observed in the southern hemisphere, which is also the location of the south polar sea – the largest and shallowest confirmed reservoir of liquid water (Iess et al., 2014). Hence, it is straightforward to conclude that proximity to the liquid reservoir is an additional control on plume eruptions at Enceladus. Based on the available data, the only known liquid water reservoir on Europa is its global ocean. If the global ocean is the source of plume material at Europa, it should be accessible to fractures wherever they are on Europa. To restrict plume activity to only two plumes, detectable in only one observation, would require that either tidally-active fractures or fractures that penetrate all the way to the ocean are rare. Otherwise, as stated above, there should be a plethora of plumes. Given the challenges in opening a fracture at depth with tensile tidal stress, which is swamped out by overburden stress less than a km from the surface, limited access to the ocean may be a reasonable conclusion.

Some models of the formation of chaotic terrains (e.g. Schmidt et al., 2011; Michaut and Manga, 2014) and ridges along fractures (e.g. Greenberg et al., 1998; Dombard et al., 2013; Johnston and Montesi, 2014) appeal to liquid water within Europa's ice shell. Although unconfirmed, these smaller, shallower reservoirs are more appealing plume sources than the global ocean because they would be easier to access with low tidal stresses and only those fractures that interact with a reservoir would generate plumes. Hence, there is a natural way of limiting plume generation to particular, but otherwise ordinary, fractures. A shallow reservoir may also provide a mechanism for creating short-lived plumes.

In our analysis, we have interpreted the non-detections reported by Roth et al. (2014a,b) as a lack of plume eruptions. However, the non-detections are also consistent with a reduction in plume output of at least a factor of 2 to 4, depending on the particular observation. A continuously active plume that varies in eruptive output would be similar to the plume behavior observed at Enceladus. However, Enceladus' plume varies on its orbital timescale, with peaks in emission repeatedly occurring at similar orbital phase. The fact that Europa's emission can vary at the same time in its orbit points to an additional mechanism that alters the stress state on the source fractures over a longer timescale (e.g. precession) or short-lived plumes. Tidal models that include stress solely from eccentricity or that assume minimal spin pole precession across all of the HST observations would still be incompatible with a long-lived, tidally-controlled plume, even if the non-detections indicate less emission rather than no emission. It is worth noting that detections can also be hampered by local plasma conditions, although we prefer not to invoke a false-negative detection to explain the drop in emission in the Jan 2014 observations.

Based on our analysis, we have identified several key differences between plumes on Europa and Enceladus.

(1) The Enceladean plumes are long-lived relative to the tidal cycle; the plumes provide the source material for Saturn's E-ring (Kempf et al., 2010), which has been observed for ~50 years (Baum et al., 1981 and references therein). The lifetime of the plumes observed on Europa is not yet known, but the lack of detections in the *Voyager* and *Galileo* data sets (Phillips et al., 2000) suggests that Europa's plumes are not as long-lived as those on Enceladus. This may be related to a difference in liquid reservoirs – the regional sea on Enceladus versus shallow pockets of water within Europa's shell – or perhaps viewing geometry.

(2) On Enceladus, the majority of plume material clearly emanates from the Tiger Stripe fractures and has been sourced from that region for at least the decade during which *Cassini* has observed Enceladus (Porco et al., 2006; Spitale and Porco, 2007; Porco et al., 2014; Spitale et al., 2014, submitted). On Europa, we have not identified analogous source fractures within the current imaging data set, which may be a result of poor image coverage. However, coupled with the lack of long-lived eruptions, another plausible explanation is that the source regions of tidally-controlled plumes on Europa vary with time. In that case, we should not expect the same source faults to consistently produce plumes, perhaps leading to less distinctive source fractures.

(3) The source fractures for the Enceladean plumes are oriented such that the majority of their lengths experience tensile tidal stress when Enceladus is at apocenter. Europa's plumes were also detected near apocenter, which was initially interpreted as evidence that, like at Enceladus, the plumes were related to tides. However, the timing of tensile stress on fractures depends on their orientations as well as their locations. If the Tiger Stripe fractures were oriented differently, they would not necessarily be in tension at apocenter. In fact, there is some variability in the orientations of the Tiger Stripe fractures along their lengths, which causes individual sections to be in tension at different points in the orbit. The observed temporal variation in plume intensity may be related to the percentage of the Tiger Stripes in tension at a given time in the orbit (Hurford et al., 2014; Nimmo et al., 2014). The source fractures for the plumes on Europa cannot experience peak tidal tensile stress at apocenter (180°) and be consistent with the non-detections. Rather, for the candidate fractures we identify for Europa, the tensile stress is approaching zero near apocenter, and the peak tidal tensile stress occurs when Europa is near 120°. Therefore, observing Europa when it is at apocenter will not guarantee plume detections, nor can non-detections at apocenter rule out tidally-controlled plumes, because the orientations of the source fractures are unknown.

6. Conclusions

The detection of water vapor plumes on Europa is exciting because it suggests that liquid water may reach the surface, with important astrobiological implications, and because it is the first direct evidence of present-day geologic activity on Europa. Further observations and characterization of plumes is certainly a worthy endeavor, and any future mission to Europa should be equipped to do so. However, based on our analysis, we expect that the locations and timing of large-scale plume activity on Europa will likely be much harder to predict than at Enceladus because the plumes may not be long-lived and because the source locations may change over time (see Section 5.3). Hence, the focus of our exploration should be to identify and characterize all plumes, not solely to study the specific plumes identified by Roth et al. (2014a). Furthermore, there are significant differences in the characteristics of plumes on Europa and Enceladus, as detailed in Section 5.3. Any plans for future exploration of Europa's plumes should carefully consider these differences and not be tailored to the behavior observed at Enceladus.

We have assessed the possible conditions under which the observed plumes on Europa could be tidally-controlled. We find it very unlikely that tidal stress only from eccentricity or with a constant spin pole direction would produce the observed difference in plume activity when Europa was at similar orbital positions. However, if the spin pole precessed by at least 60° between the Dec 2012 and Jan 2014 observations, we do find a more plausible set of candidate faults that are consistent with the observations. The precession rates implied by this model are well within the expectations from gravitational models but perhaps faster than the rate inferred from tidal-tectonic modeling of lineaments and strike-slip faults. A key test of this model would be to observe Europa at an orbital position near 120°, when tensile stress on the candidate fractures would be near its peak. A non-detection at that time would imply that the plumes are either not tidally controlled or that the timescale of activity is short relative to the time period of the observations (i.e. years to decades).

Even if the plumes are never again detected, or the original detection is shown to be erroneous, this work has demonstrated that obliquity and spin pole precession can augment the tidal stress field enough to change the plume behavior we would predict. Hence, future endeavors to identify or study plumes should account for the effects of Europa's obliquity.

The locations and azimuths of candidate fractures are very similar for all precession rates we tested and whether we use the thin shell or viscoelastic tidal stress equations. However, the particular set of candidate fractures would likely change with more sophisticated assumptions about the mechanics of plume eruptions. The candidate faults we identify are ordinary in terms of the stress they experience. There would be no clear reason to select these faults a priori; it is only the pattern of observed detections that cause us to select these particular locations and azimuths. Comparing the candidate faults from our tidal model with images of Europa's surface reveals no obvious source fractures, but the lack of global imagery at a suitable resolution for mapping is certainly a limiting factor in identifying plume sources.

A large number of faults are in tension during observations that did not detect large-scale plumes, which suggests that tidal stress is not the only mechanism controlling eruptions. Limited access to liquid water likely plays a key role. If the global ocean is the source of plume material, then perhaps few fractures are able to penetrate through the entire ice shell, limiting eruptions. Alternatively, the plume source material may come from shallow reservoirs within the shell. In that case, the spatial and temporal distribution of water within the ice shell should exert an additional control on plume activity as well as the number and distribution of active fractures.

Continued observation and analysis of plumes on both Europa and Enceladus can provide insight into the subsurface structure, the fracture and fluid transport processes, and the distribution and accessibility of liquid water within the ice shells of these moons.

Further characterization of plumes and surface geology, along with ground-based measurements of Europa's rotation state, will enhance our understanding of Europa's dynamical and thermal evolution and the relationship between tides and geologic activity.

Acknowledgements: A. Rhoden was partially supported through an appointment to the NASA Postdoctoral Program, administered by ORAU. L. Roth and K. Retherford were supported through HST Program number 13619 provided by NASA through a grant from the Space Telescope Science Institute, which is operated by the Association of Universities for Research in Astronomy, Inc., under NASA contract NAS5-26555.

Table 1: Summary of HST observations from Roth et al. (2014a,b)

| Date | Start time | True anomaly range, total | Mid-point time | True anomaly, mid-point time | Detection | Days since previous observation |
|---|---|---|---|---|---|---|
| 1999-10-05 | 8:39 | 343-13 | 11:56 | 357 | No | - |
| 2012-11-08 | 20:41 | 286-316* | 23:59 | 301* | No | 4783 |
| 2012-12-30 | 18:49 | 189-218* | 22:06 | 203* | Yes | 52 |
| 2014-01-22 | 14:02 | 191-221 | 17:19 | 205 | No | 388 |
| 2014-02-02 | 8:20 | 208-236 | 11:33 | 222 | No | 11 |

* Values are updated and differ by ~2º from the values reported in Roth et al. 2014a.

Table 2: Summary of precession model results

| Dec 2012 SPD | Jan 2014 SPD | Δ SPD | Precession rate | Number of candidate fractures | Max tensile stress in Dec 2012 |
|---|---|---|---|---|---|
| 270 | 90 | 180 | 0.464 | 676 | 15.8 kPa |
| 315 | 45 | 90 | 0.232 | 553 | 11.3 kPa |
| 270 | 0 | 90 | 0.232 | 548 | 12.0 kPa |
| 330 | 30 | 60 | 0.155 | 471 | 8.6 kPa |
| 270 | 330 | 60 | 0.155 | 419 | 2.8 kPa |

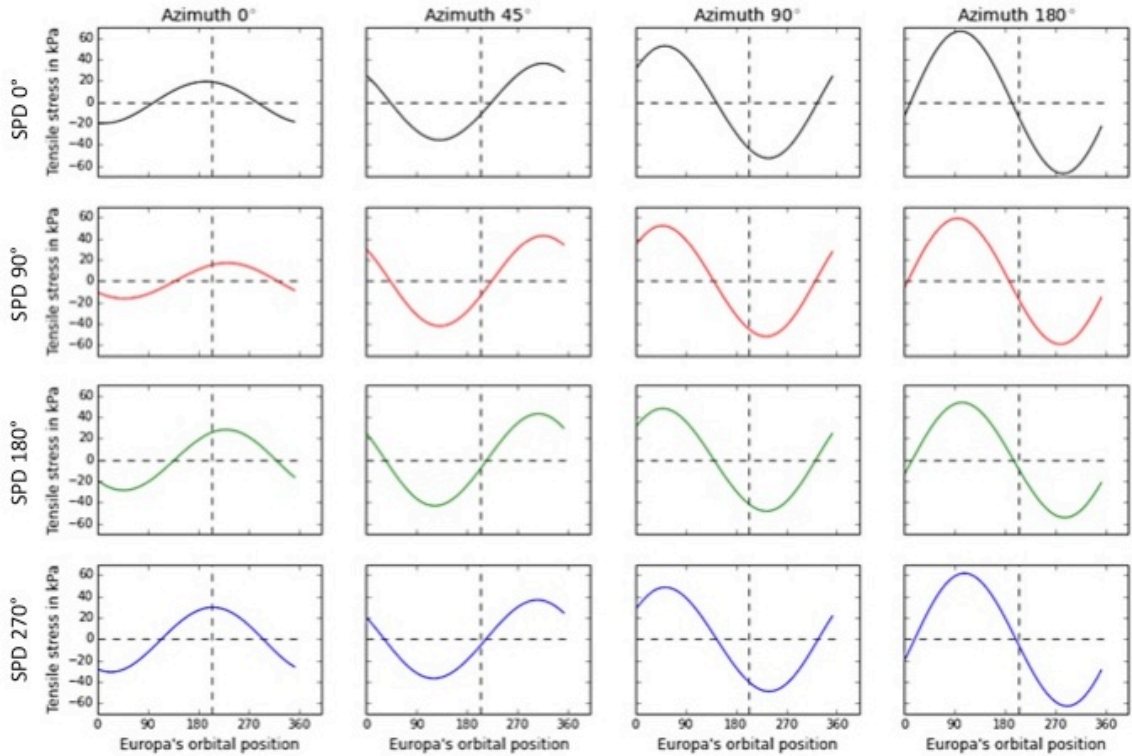

Figure 1: Tidal normal stress on hypothetical faults at four different azimuths (columns) and at four different spin pole directions (rows). The horizontal dashes lines show the transition between tensile (+) and compressive (-) stress. The vertical dashed lines show Europa's orbital time when the plume was detected. A non-detection was also reported when Europa was at nearly the same orbital position. As shown here, the maximum tensile stress on a fault changes in magnitude over one spin pole precession period. For a 0° fracture (N-S trending; left-most column), the maximum stress is always tensile, while the stresses on 45° and 90° faults (middle columns) are always compressive. Only the 180° fault experiences both tension and compression during one precession period, perhaps explaining the detection and non-detection at the same time in Europa's orbit. For any given region, there are a range of azimuths that experience both tension and compression throughout a precession period.

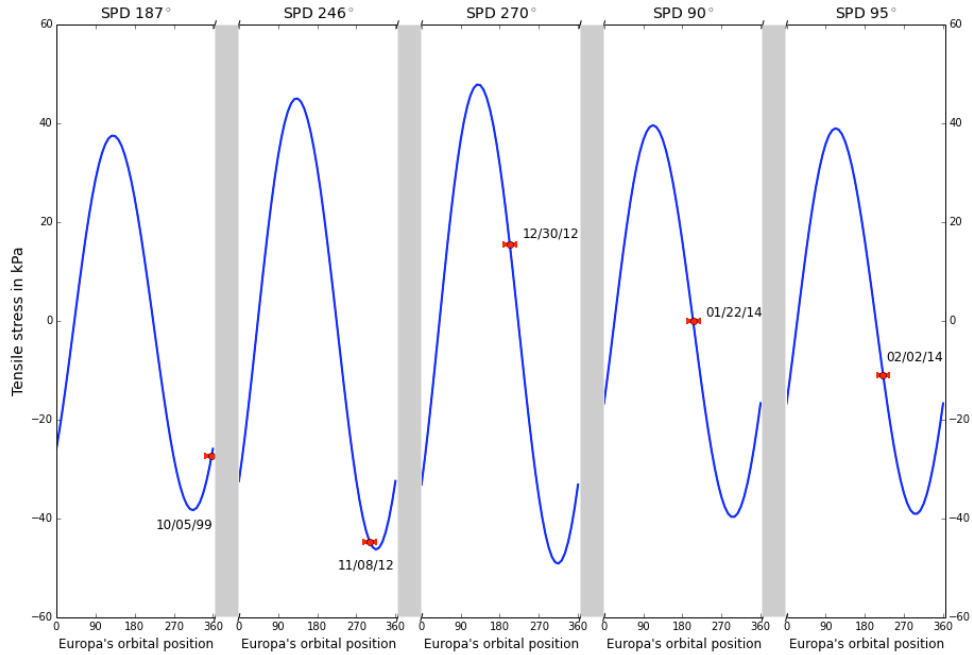

Figure 2: Tidal normal stress over multiple orbits (blue line), assuming a non-zero obliquity and rapidly precessing spin pole, on a fault that experiences tension only during the Dec 2012 observation. Precession causes the SPD to change between observations leading to the different stress curves that are separated by the grey bars in this figure. We assume a spin pole direction of 270° for the Dec 2012 observation and an SPD of 90° for the Jan 2014 observation. By applying a constant precession rate, we derive SPDs for the other observation times. Stress throughout one orbit is shown for each observing window, which corresponds to its specific SPD (as labeled). Red, filled symbols mark Europa's orbital position at the midpoint of each observing window, while the red line indicates the full range of positions covered during each 7-hr window. In this model, it is the change in spin pole direction, rather than Europa's orbital position, that allows the fault to be in tension in Dec 2012 and compression in Jan 2014, despite the overlap in Europa's orbital position.

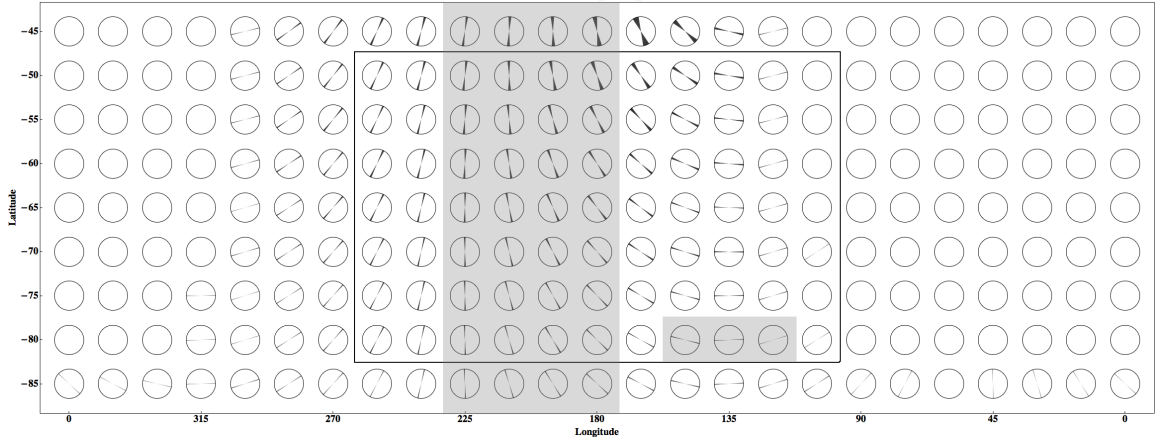

Figure 3: The distribution of candidate faults (black wedges) using the tidal model that produces the largest number of candidate faults, in which we assume an obliquity of 0.1°, an SPD of 270° in Dec 2012, and an SPD of 90° in Jan 2014. The black outline represents the most likely location of the candidate faults, according to Roth et al. (2014a,b). The grey shaded regions mark the locations of *Galileo* imaging data of ~230 m/pix resolution or better.

SOM figures and captions:

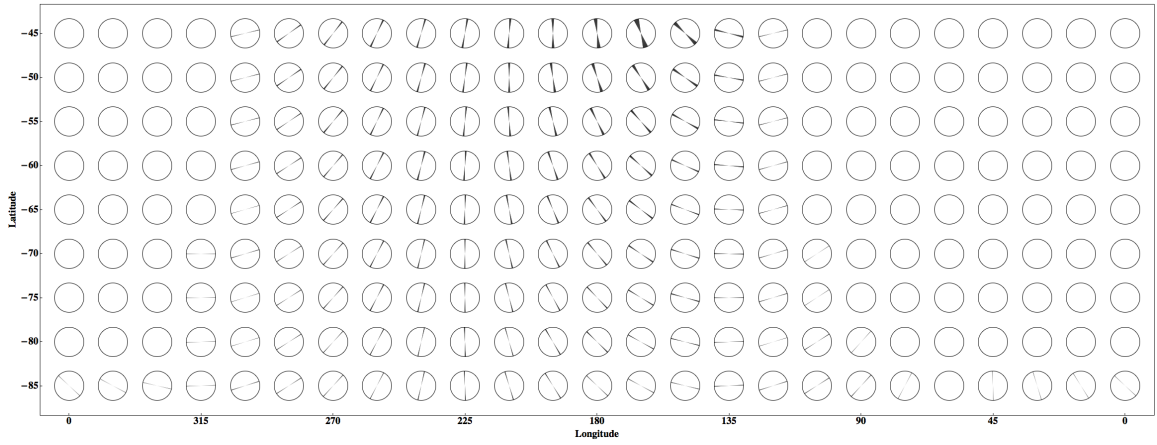

Figure 1: Candidate faults (black wedges) assuming an SPD of 315° in Dec 2012 and an SPD of 45° in Jan 2014.

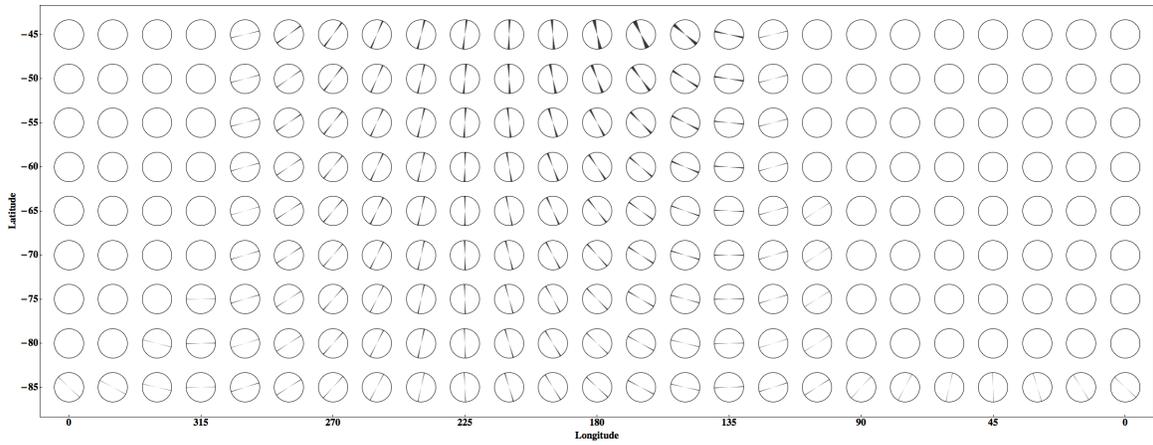

Figure 2: Candidate faults assuming an SPD of 270° in Dec 2012 and an SPD of 0° in Jan 2014.

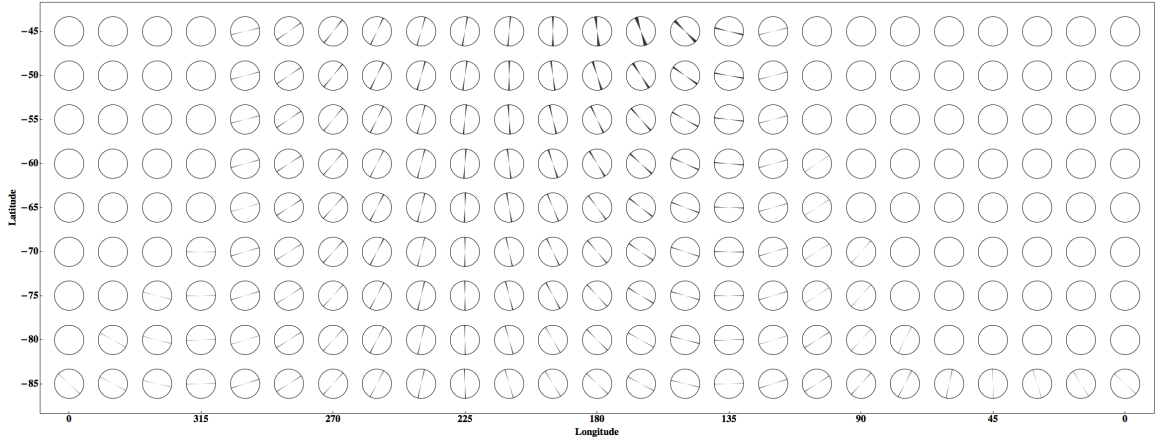

Figure 3: Candidate faults (black wedges) assuming an SPD of 330° in Dec 2012 and an SPD of 30° in Jan 2014.

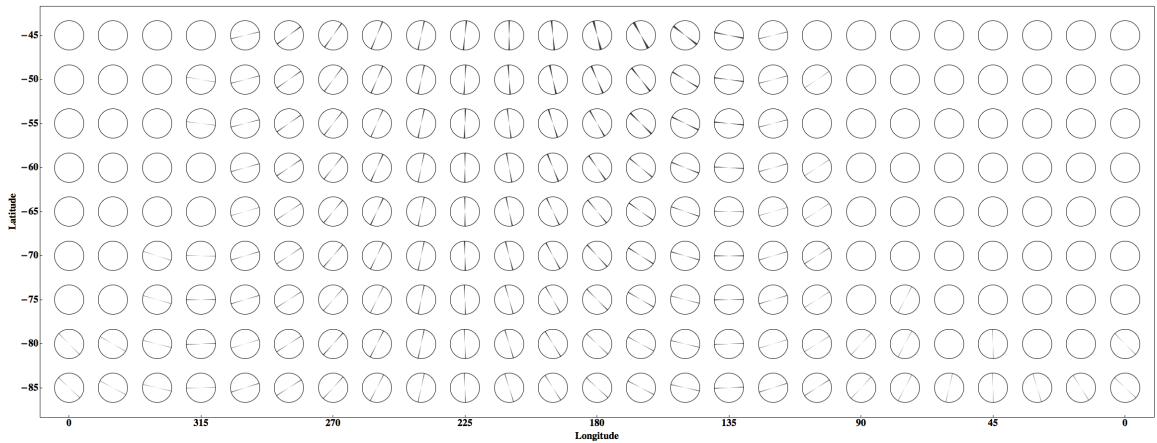

Figure 4: Candidate faults (black wedges) assuming an SPD of 270° in Dec 2012 and an SPD of 330° in Jan 2014.

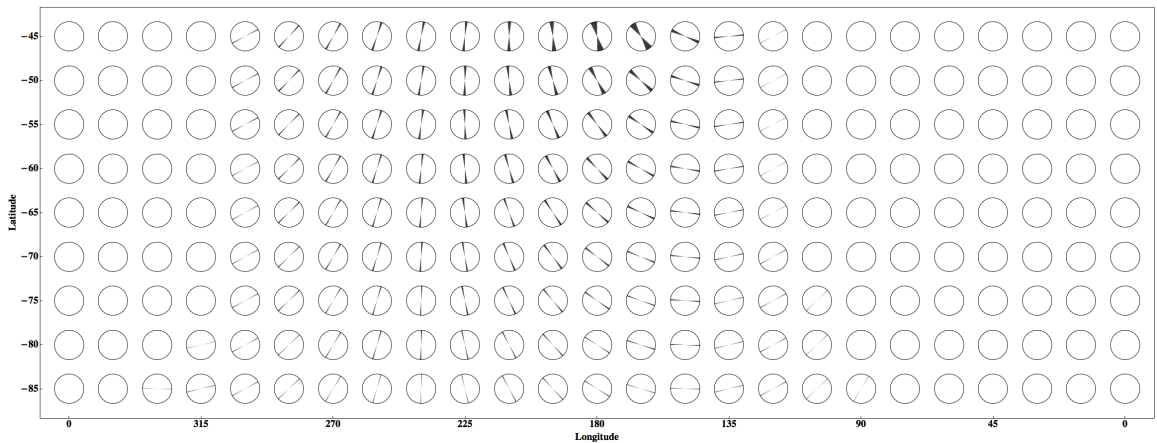

Figure 5: Candidate faults (black wedges) assuming an SPD of 270° in Dec 2012 and an SPD of 90° in Jan 2014 using the starting times of the HST observation windows rather than the midpoints. Comparing with Fig. 3 in the main text shows a small difference in the azimuths of the candidate faults depending on the time at which we make our calculations.